# Indoor Channel Measurements and Communications System Design at 60 GHz

*L. Rakotondrainibe[1], G. Zaharia[1], G. El Zein[1], Y. Lostanlen[2]*

[1]IETR-UMR CNRS 6164, 20 Av. des Buttes de Coësmes, CS 14315, 35043 Rennes Cedex, France

[2]SIRADEL, 3, Allée Adolphe Bobierre, CS 24343, 35043 Rennes Cedex-France

lrakoton@insa-rennes.fr; gheorghe.zaharia@insa-rennes.fr; ghais.el-zein@insa-rennes.fr; yves.lostanlen@ieee.org

## Abstract

This paper presents a brief overview of several studies concerning the indoor wireless communications at 60 GHz performed by the IETR. The characterization and the modeling of the radio propagation channel are based on several measurement campaigns realized with the channel sounder developed at IETR. Some typical residential environments were also simulated by ray tracing and Gaussian Beam Tracking. The obtained results show a good agreement with the similar experimental results. Currently, the IETR is developing a high data rate wireless communication system operating at 60 GHz. The single-carrier architecture of this system is also presented.

## 1. Introduction

During the last decade, substantial knowledge about the 60-GHz millimeter-wave (MMW) channel has been accumulated and different architectures have been analyzed to develop MMW communication systems for commercial applications [1-2]. The 60 GHz bandwidth is suitable for high data-rate and short-distance wireless communications. This interest is particularly due to the large bandwidth and the important power loss caused by the free space and walls attenuation which permits to reuse the same frequency bandwidth even in the next floor of the same building. Concerning the 60 GHz front-end technology, higher frequencies lead to smaller sizes of RF components including very small antennas. The cost is mainly related to the transceiver RF front ends.

The development of new wireless communication systems requires accurate knowledge of the propagation channel to efficiently simulate and design them, including new modulation schemes, coded and multiple access techniques. The rest of this paper is organized as follows: section 2 presents an overview of several studies realized at IETR concerning the measurements and characterization of the 60 GHz radio propagation channel. Section 3 reports recent work concerning of a 60-GHz radio communication system. Some conclusions are drawn in section 4.

## 2. Channel measurements and characterization

During the last decade, several research activities were carried out at IETR in the 60 GHz bandwidth: the realization of the channel sounder, the indoor radio channel measurements, simulation and characterization.

### 2.1 Channel sounder

A 60-GHz wideband channel sounder was developed at IETR (Fig. 1). This channel sounder has 500 MHz bandwidth, 40 dB relative dynamic and 2.3 ns effective time resolution, which means that two paths separated from by 70 cm can be correctly discriminated. Based on the sliding correlation technique, this sounder is optimized to perform long term measurement campaigns. Some measurement results with Doppler analysis up to 20 kHz are presented in [3].

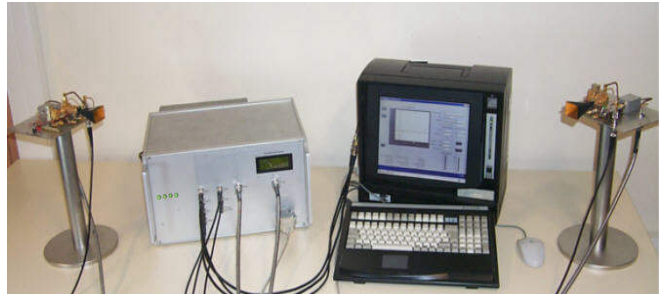

Fig. 1 Channel sounder at 60 GHz realized by IETR

## 2.2 Channel measurements and characterization

In [4-6], the results of several studies concerning the radio propagation at 60 GHz in residential environments were published. These studies are based on several measurement campaigns realized with the IETR channel sounder. The measurements have been performed in residential furnished environments. The study of the angles-of-arrival (AOA) shows the importance of openings (such as doors, staircase, etc.) for the radio propagation between adjacent rooms (Fig. 2). From the database of impulse responses, several propagation characteristics are computed: attenuation, delay spread, delay window, coherence bandwidth [5]. The wave propagation depends on antennas (beam-width, gain and polarization), physical environment (furniture, materials) and human activity. A particular attention is paid to the influence of the human activity on radio propagation. In [6], it is shown that people movements can make the propagation channel unavailable during about one second (Fig. 3).

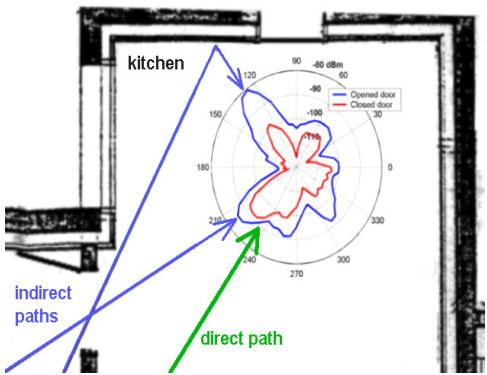
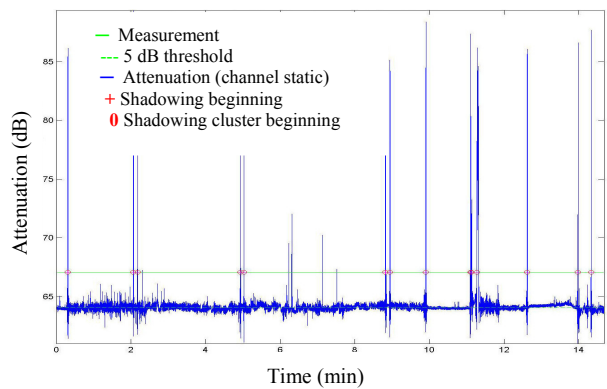

Fig. 2 Received power in the horizontal plane  
(NLOS, with a receiving horn antenna)

Fig. 3 Human activity measurement at 60 GHz  
(Receiver antenna: horn, channel activity: 4 persons)

However, the angular diversity can be used: when a path is shadowed, another one, coming from another direction, can maintain the radio link. From the characterization of the indoor radio propagation, several recommendations concerning the deployment of the very high data rate 60 GHz wireless networks are derived in [5].

## 2.3 Deterministic simulation of the 60 GHz radio channel

Two deterministic simulation tools have been used to complement the experimental characterization: a ray-tracing tool [6] and a 3 D Gaussian Beam Tracking (GBT) technique [7]. Both tools provided comparable coverage simulations in a residential indoor environment (a house) at 60 GHz and 500 MHz bandwidth. In Fig. 4, simulated coverage results based on GBT algorithm are shown. This method based on Gabor frame approach is particularly well suited to high frequencies and permits a collective treatment of rays which offers significant computation time efficiency. The advanced deterministic prediction tool X-Siradif based on ray-tracing has further shown a good agreement with the power-delay measurement results (see Fig. 5), which is important for such a wide bandwidth. Besides the ray-tracing technique can easily and accurately take into account any measured antenna patterns. In [6], other comparison results concerning the channel impulse response and the angles-of-arrival are given. Besides, the ray-tracing technique can easily and accurately take into account any measured antenna patterns.

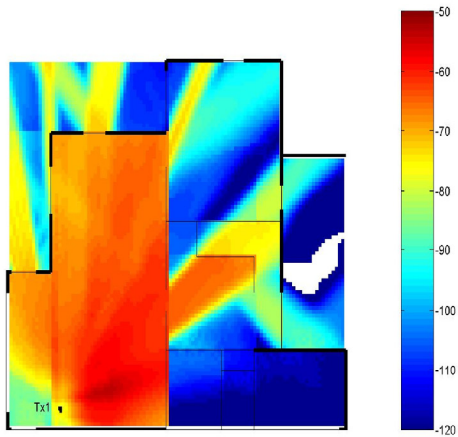 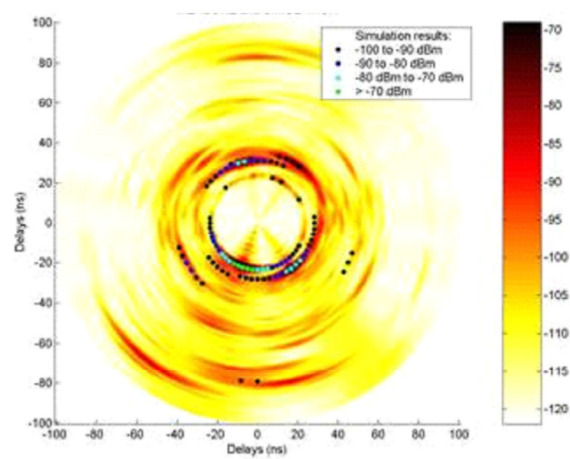

Fig. 4 Power coverage (GBT algorithm)    Fig. 5 Measurement and simulation of the channel impulse response

However at 60 GHz static simulations are not sufficient to help for the design of communication systems. SIRADEL is still pursuing its investigation on indoor propagation by setting up dynamic scenarios to simulate the presence of people in the room (mobility models, direct path and obstruction).

## 3. Design of a 60 GHz wireless communication system

Studies for different types of single carrier (SC) modulation and orthogonal frequency division multiplexing (OFDM) were proposed in the IEEE 802.15.3 Task Group 3c for the future wireless communication systems operating at 60 GHz. The choice of modulation schemes for 60 GHz radio is highly dependent on the characteristics of the propagation channel, the use of high gain antenna/antenna array and the limitations imposed by the RF technology. Some SC modulations as BPSK and QPSK are considered. The SC has two main advantages over OFDM: lower complexity and lower PAPR (peak-to average power ratio).

IETR, one of the partners of the Techim@ges project, is currently involved in the design and the realization of a low-cost, high data rate (about 1 Gbps) and small-distance (d < 10 m) wireless communication system. Fig. 6 shows the single carrier system architecture proposed by IETR which uses a BPSK modulation. The 3.5 GHz and 58.5 GHz phase locked oscillators use a 70 MHz frequency synthesizer. At the receiver, the intermediate frequency and the clock are obtained from the received signal. The frequency-domain equalizer (FDE) and some error correcting codes are under study. In order to improve the link budget, especially for point-to-point applications, it is preferable to use high-gain directive antennas.

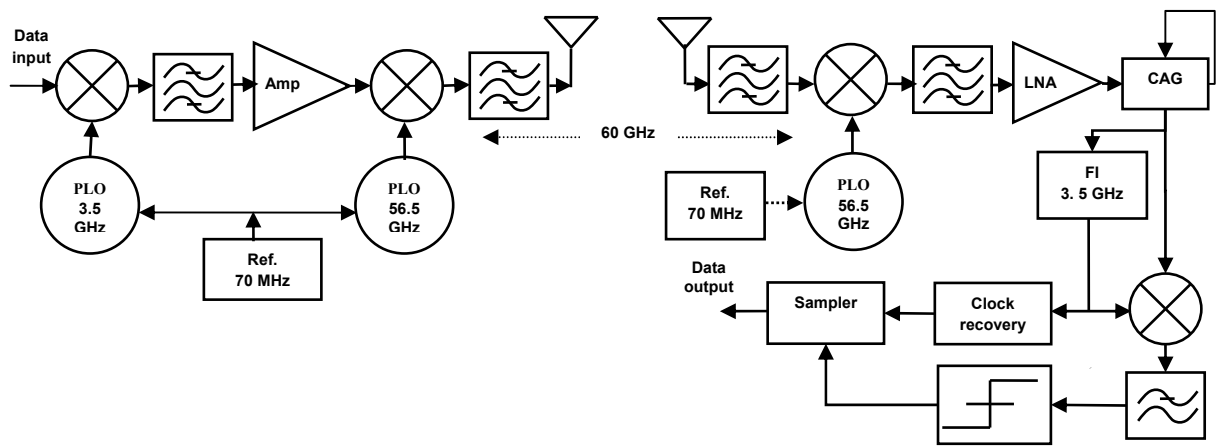

Fig. 6 Single-carrier wireless communication system at 60 GHz

## 4. Conclusion

In this paper, a brief overview of several studies performed at IETR on 60 GHz indoor wireless communications is presented. The characterization of the radio propagation channel is based on several measurement campaigns realized with the channel sounder of IETR. Some typical residential environments were also simulated by ray tracing and Gaussian Beam Tracking. The obtained results show a good agreement with the experimental results. Currently, the IETR is developing a SC wireless communication system operating at 60 GHz.

## 5. Acknowledgments

The study on the 60 GHz indoor radio channel was supported by the French National Research Network in Telecommunications RNRT COMMINDOR. The realization of the 60 GHz wireless communications system is part of the research project Techim@ges supported by the French "Media & Network Cluster" and the COMMIDOM project of the "Région Bretagne".

## 6. References

1. H. Yang, P. F. M. Smulders, and M. H. A. J. Herben, "Channel Characteristics and Transmission Performance for Various Channel Configurations at 60 GHz", *EURASIP Journal on Wireless Communications and Networking*, Volume 2007, ID 19613, 15 pages, March 2007.
2. N. Guo, R. C. Qiu, S. S. Mo, and K. Takahashi, "60-GHz Millimeter-Wave Radio: Principle, Technology, and New Results", *EURASIP Journal on Wireless Communications and Networking*, ID 68253, 8 pages, Sept. 2006.
3. S. Guillouard, G. El Zein, and J. Citerne, "Wideband Propagation Measurements and Doppler Analysis for the 60 GHz Indoor Channel", *IEEE MTT-S Digest*, 1999.
4. S. Collonge, G. Zaharia, and G. El Zein, "Wideband and Dynamic Characterization of the 60 GHz Indoor Radio Propagation-future Home WLAN Architectures", *Annals of Telecommunications*, special issue on WLAN, March-April 2003, Vol. 58, N° 3-4, pp. 417-447.
5. S. Collonge, G. Zaharia, G. El Zein, "Influence of the human activity on wide-band characteristics of the 60 GHz indoor radio channel", *IEEE Trans. on Wireless Communications*, Vol. 3, Issue 6, Nov. 2004, pp. 2396-2406.
6. Y. Lostanlen, Y. Corre, Y. Louët, Y. Le Helloco, S. Collonge, G. Zaharia, and G. El Zein, "Comparison of Measurements and Simulations in Indoor Environments for Wireless Local Networks at 60 GHz", *IEEE Vehicular Technology Conference 2002*, Birmingham, USA, May 2002.
7. R. Tahri, D. Fournier, S. Collonge, G. Zaharia, and G. El Zein, "Efficient and Fast Gaussian Beam-Tracking Approach for Indoor-Propagation Modelling", *Microwave and Optical Technology Letters*, Vol. 45, N° 5, June 2005, pp. 378-381.